\documentclass[12pt]{article}
\newcommand\be{\begin{equation}}
\newcommand\ee{\end{equation}}
\newcommand\ba{\begin{eqnarray}}
\newcommand\ea{\end{eqnarray}}
\newcommand\eq{\begin{equation}}           
\newcommand\en{\end{equation}}

\input epsf
\usepackage{amssymb,amsmath,graphicx} 
 
\begin{document}

\title{
{\hfill \small UMN-TH-2540/07 \\
\hfill \small FTPI-MINN-07/07
\\ ~\\~\\}
Warped Leptogenesis with Dirac Neutrino Masses
}
\author{Tony Gherghetta$^1$, Kenji Kadota$^{1,2}$ and Masahide Yamaguchi$^3$
\\
\\
{\em \small $^1$School of Physics and Astronomy, University of Minnesota, Minneapolis, MN 55455}\\
{\em  \small $^2$ William I. Fine Theoretical Physics Institute, University of Minnesota, Minneapolis, MN 55455}\\
{\em \small $^3$ Department of Physics and Mathematics, Aoyama Gakuin University, Sagamihara 229-8558, Japan}}
\maketitle   
\begin{abstract}
We show how leptogenesis can occur at the TeV scale 
with neutrinos that possess almost purely Dirac masses and negligible Majorana 
mass contributions as a consequence of the small wavefunction overlap in a warped fifth dimension. 
Lepton number violation at the Planck scale is introduced via a Majorana mass term on the Planck brane. Such a Majorana mass operator leads to the small mass splitting of otherwise 
degenerate Kaluza-Klein excited states on the TeV brane. This tiny mass splitting can 
compensate for the small Yukawa couplings to give a CP asymmetry large enough to 
produce the sufficient baryon asymmetry from the decay of the nearly degenerate neutrino
Kaluza-Klein  states. In this way the standard baryogenesis via leptogenesis scenario can naturally occur at the TeV scale without the need for a high mass scale.
\end{abstract}

\setcounter{footnote}{0} 
\setcounter{page}{1}
\setcounter{section}{0} \setcounter{subsection}{0}
\setcounter{subsubsection}{0}

\newpage
\section{Introduction}
The generation of a baryon asymmetry of the Universe is a fundamental
question in particle physics and cosmology. The usual baryogenesis via 
leptogenesis scenario \cite{lepto} occurs when a right-handed neutrino decays 
out of equilibrium and then sphalerons reprocess the lepton asymmetry into a baryon 
asymmetry above the electroweak scale. This is a theoretically appealing possibility, 
except that the leptogenesis scenarios involving heavy right-handed neutrinos 
needed for the see-saw mechanism \cite{seesaw} are difficult to experimentally verify. 
In addition to the difficulties in probing the nature of heavy right-handed
neutrinos via energetically accessible light (left-handed) neutrinos,
keeping track of the history of the universe from such a high energy
scale, (e.g. GUT scale), down to the present time is not a
trivial issue and it would be desirable if the observed baryon asymmetry
was produced at a much lower energy scale where it could be directly accessible by
terrestrial or astrophysical experiments.

We will show that the usual leptogenesis scenario can in fact be implemented at the TeV scale 
with predominantly Dirac neutrino masses and a small lepton number violating Majorana
contribution. All small parameters naturally arise from the warped geometry. In particular 
the tiny Dirac Yukawa couplings result from the small 
wavefunction overlap of the fields in the warped fifth dimension. 
Interestingly, even though the neutrino is essentially Dirac in nature, 
there can still be sufficient lepton number violating effects to realize the leptogenesis. This is done
by including a Majorana mass term on the Planck brane which represents the expected breaking of global symmetry by the higher dimensional operators induced from Planck scale physics. The right-handed neutrinos are localized near the TeV brane, so that the effects of the UV-localized Majorana mass operator 
on the standard model neutrino masses, are highly suppressed at the TeV brane realizing the Dirac neutrino masses. The Standard Model matter fields are localized throughout the bulk to obtain the necessary Yukawa couplings via wavefunction overlap with a Higgs field near the TeV brane~\cite{fermass4}\footnote{If the right handed neutrinos are instead localized near
the Planck brane, the Majorana mass contributions become dominant
leading to the see-saw mechanism in the warped extra dimensions
\cite{majora}. Alternatively, if one imposes lepton number conservation even on the Planck
brane, then another possibility is a Dirac neutrino mass scenario which localizes
the right (left) handed neutrino near the Planck (TeV) brane~\cite{fermass1}.}.

Since the CP asymmetry due to the right-handed neutrino decay is proportional to the Yukawa couplings squared, it may seem that the resultant baryon asymmetry is too small to account for the current baryon asymmetry of the universe. However, as we shall point out, the baryon asymmetry is also inversely proportional to the small mass difference of the almost  degenerate Kaluza-Klein Majorana states.
Consequently, the Majorana mass term localized on the Planck brane gives a desirable tiny mass splitting between the even and odd excited Kaluza-Klein (KK) modes to compensate for the small Yukawa couplings. Thus the requisite lepton asymmetry is generated
which is then reprocessed into a baryon asymmetry by electroweak sphalerons.
Furthermore, in the four-dimensional (4D) interpretation of our model, the lowest lying 
Kaluza-Klein states are composite and are thermally produced at the TeV scale. 
Hence our scenario is experimentally verifiable because the Kaluza-Klein states could be directly produced at the LHC.

After introducing the setup and notation of the warped extra-dimension model in \S~\ref{set}, we discuss the properties of the bulk neutrinos in the presence of boundary Majorana masses in \S ~\ref{majprop}. In \S~\ref{lep} we present analytical estimates for the parameter constraints to produce the desirable baryon asymmetry of the Universe. This includes constraints arising from the electron and electron-neutrino Yukawa couplings in the standard model that are consistent with experimental observations. Finally our discussion/conclusion is in  \S ~\ref{dis}.

\section{Setup}
\label{set}
Consider the fifth dimension $y$ compactified on an orbifold $S^1/Z_2$ 
of radius $R$, with $-\pi R\leq y \leq \pi R$, which is bounded by two three-branes at the orbifold fixed points $y=0, \pi R$ known as the UV (or Planck) and IR (or TeV) brane respectively. The five-dimensional (5D) Einstein's equations for this geometry lead to \cite{rs}
\ba
\label{adsmet}
ds^2=e^{-2k y}\eta_{\mu\nu}dx^{\mu}dx^{\nu}-dy^2~,
\ea
where the AdS curvature radius is $1/k$ and 4D metric is $\eta_{\mu \nu}=\mbox{diag}(1,-1,-1,-1)$.

The bulk action for the 5D Dirac spinor, consisting of two two-component spinors 
$\Psi=(\psi,\bar{\chi})^T$, has the terms~\cite{fermass1,fermass2}
\begin{eqnarray}
\int d^5 x  \sqrt{-g}
\left[ -i\bar{\psi}\bar{\sigma}^{\mu}\partial_{\mu} \psi
-i\bar{\chi}\bar{\sigma}^{\mu}\partial_{\mu}{\chi}
+\frac12
(\chi \overleftrightarrow{\partial_5}\psi-\bar{\psi}\overleftrightarrow{\partial_5}   \bar{\chi})
\right.\nonumber\\
\qquad\qquad \left.
+m_D(\chi\psi+\bar{\psi}\bar{\chi})
+\frac12 m_M (\psi \psi+\bar{\chi}\bar{\chi}+h.c.)\right]~,
\end{eqnarray}
where $\overleftrightarrow{\partial_5} = \overrightarrow{\partial_5}-\overleftarrow{\partial_5}$. The term $\psi \partial_5 \chi$, for example, 
implies that $\psi \chi$ is odd under the $Z_2$ action $y\rightarrow -y$ and, for definiteness, we assign an even parity for $\psi$ and an odd parity for $\chi$ in the rest of this section. 
We parametrize this bulk Dirac mass, which should be odd under $Z_2$, in terms of the step function 
$\epsilon(y)\equiv y/|y|$
\ba
m_D=c~k\epsilon(y)~,
\ea
where $c$ is a dimensionless parameter.
We are also interested in the right-handed neutrino Majorana mass completely localized on the Planck brane which can be parameterized as
\ba
m_M=d_M\delta(y)~,
\ea
where $d_M$ is a dimensionless constant.
This is motivated from the fact that global symmetries are in general expected to be broken by Planck-scale physics. Hence a Majorana mass term can well be located on the Planck brane, which leads to negligible lepton-number violation on the TeV brane.

In five dimensions fermion fields can be decomposed as
\ba
\psi(x,y)=\frac{e^{2k y}}{\sqrt{2\pi R}}
\sum_{n=0}^\infty\psi^{(n)}(x)f^{(n)}_{+}(y), ~~
\bar{\chi}(x,y)=\frac{e^{2k y}}{\sqrt{2\pi R}}
\sum_{n=1}^\infty\bar{\chi}^{(n)}(x)f^{(n)}_{-}(y)~,
\ea
where $+(-)$ indicates an even (odd) parity under $Z_2$. Because the Majorana mass is confined on the Planck brane, the bulk equations of motion have the same form as those without the boundary Majorana masses and, consequently, the solutions are
\ba
\label{fsol}
{\rm Re} f_+^{(n)}(y)&=&\frac{e^{k y/2}}{N_n}\left[J_{|c+1/2|}\left(\frac{m_n}{k e^{-ky}}\right)
-\frac {J_{|c+1/2|\pm1}\left(\frac{m_n}{ke^{-\pi k R}}\right)}
{Y_{|c+1/2|\pm1}\left(\frac{m_n}{ke^{-\pi k R}}\right)}
Y_{|c+1/2|}\left(\frac{m_n}{k e^{-ky}}\right)\right]~,
\nonumber
\\
{\rm Re}f_-^{(n)}(y)&=&\frac{e^{k y/2}}{N_n}\left[J_{|c-1/2|}\left(\frac{m_n}{k e^{-ky}}\right)
-\frac {J_{|c-1/2|}\left(\frac{m_n}{ke^{-\pi k R}}\right)}
{Y_{|c-1/2|}\left(\frac{m_n}{ke^{-\pi k R}}\right)}
Y_{|c-1/2|}\left(\frac{m_n}{k e^{-ky}}\right)\right]~,
\ea
with $c<-1/2\,(c>-1/2)$ for the even parity field and the normalization constants $N_n$ obtained from  
\ba
\frac{1}{2\pi R}\int^{\pi R}_{-\pi R}
dy~e^{k y}f^{(n)*}_\pm(y)f^{(m)}_\pm(y)=\delta_{nm}~.
\ea 
The spectrum of KK masses $m_n$ are determined by the boundary conditions (we choose the basis such that $d_M$ is real)
\ba
\label{majoranabc}
{\rm Re} f_-^{(n)}(0)-\frac{d_M}{2} {\rm Re} f_+^{(n)}(0)=0~, \nonumber \\
{\rm Re} f_-^{(n)}(\pi R)=0~.
\ea
The solutions for the imaginary parts of $f^{(n)}_{\pm}$ can be obtained by switching the sign of the Majorana mass in the boundary condition.

The analogous procedures can be applied for the KK decomposition of the scalar field $\Phi$ as well whose action contains
\ba
\int d^5 x \sqrt{-g}\left(|\partial_M \Phi|^2-m_{\phi}^2|\Phi|^2\right)~.
\ea
For the KK decomposition 
\ba
\Phi(x,y)=\frac{1}{\sqrt{2\pi R}} \sum_{n=0}^\infty\Phi^{(n)}(x)f_n(y),~~
\frac{1}{2\pi R}\int^{\pi R}_{-\pi R}dy~e^{-2k y}f_n(y)f_m(y)=\delta_{mn}~,
\ea
and $m_{\phi}^2$ containing both bulk and boundary masses parameterized by
\ba
m_{\phi}^2=ak^2+2bk[\delta(y)-\delta(y-\pi R)]~,
\ea
the zero mode solution is given by
\ba
f^{(0)}(y)=\frac{e^{bky}}{N_0}~,\qquad \frac{1}{N_0}=\sqrt{\frac{2(b-1)\pi kR}{e^{2(b-1)\pi kR}-1}}~.
\ea
In the above, the boundary mass of form $b=2\pm \sqrt{4+a}$ is assumed without which 
zero-th KK mode would vanish \cite{fermass2}. 

The bulk mass parameters $c>1/2$ ($c<1/2$) and $b<1$ ($b>1$) correspond to the localizations of the zero mode wave functions around Planck (TeV) brane for Fermion and scalar fields respectively.

\section{Right-handed neutrinos in extra dimensions}
\label{majprop}
We can obtain the masses of the KK states in the presence of the Majorana mass from the boundary conditions Eq. (\ref{majoranabc}). However the solutions are not in a useful form for 
the analytical estimation of the excited KK masses, so instead we obtain the KK mass 
spectrum of the right-handed neutrino using the basis $\{{\bar f}_{\pm}^{(n)}\}$ which is obtained without including the Majorana mass \footnote{We 
can treat ${\bar f}_{\pm}^{(n)}$ to be real because the bulk equations of motions and the boundary conditions are identical for ${\bar f}_{\pm}^{(n)}$ and their conjugate ${\bar f}_{\pm}^{(n)*}$.}. In this case, the integration 
of $\{{\bar f}_{\pm}^{(n)}\}$ over the extra dimension receives the contributions from the boundary Majorana mass terms, which gives the mixing among the KK states.  
Then the diagonalization of this resultant mass matrix can give the required KK state masses as performed below
\footnote{Note that the matrix includes all the KK states 
up to some UV cutoff scale.}.
For notational clarity, we use $\{f_{\pm}^{(n)}\}$, instead of $\{{\bar f}_{\pm}^{(n)}\}$, in the following discussions to denote the  basis without the Majorana mass term. 

For the Majorana mass confined on the Planck brane, the KK states for a left-handed neutrino $\nu$ and 
those for a right-handed neutrino $N$ do not mix in the mass eigenstates before the electroweak symmetry is spontaneously broken by the finite Higgs VEV. The corresponding mass terms with the vanishing Higgs VEV for KK states of $N(x,y)=(N_+(x,y), \bar{N}_-(x,y))^T$ ($+(-)$ indicates even (odd) under $Z_2$ orbifold symmetry) are 
\ba
\label{matrixmaj}
\frac12 ({N}_+^{(0)},{N}_+^{(1)},N_-^{(1)},\ldots)
 \left( 
\begin{array}{cccc}
A_{00}&A_{01}&0&\ldots\\
A_{01}&A_{11}&D_{N1}&\ldots\\
0&D_{N1}&0&\ldots \\
\vdots& \vdots& \vdots&\ddots
\end{array}
\right)
\left(
\begin{array}{c}
{N}_+^{(0)}\\
{N}_+^{(1)}\\
N_-^{(1)}\\
\vdots
\end{array}
\right)~,
\ea
where we have used the basis of right-handed neutrinos such that the mass matrix elements are real. The spectrum of KK (Dirac) masses $D_{Nm}$ for $N$ is determined by the boundary conditions Eq. (\ref{majoranabc}) with $d_M=0$
and 
\ba
A_{mn}&=&\int ^{\pi R}_{-\pi R}\frac{dy}{2\pi R}m_M(y)f_{N+}^{(m)}(y)f_{N+}^{(n)}(y)
=\frac{1}{2\pi R}d_M f_{N+}^{(m)}(0)f_{N+}^{(n)}(0)~,\\
N_{\pm}(x,y)&=&\frac{e^{2k y}}{\sqrt{2\pi R}}
\sum_{n=0}^{\infty}N^{(n)}_{\pm}(x)f^{(n)}_{N\pm}(y)~, \\
f^{(0)}_{N+}(y)&=&\frac{e^{-c_Nk y}}{N_0},~~
\frac{1}{N_0}=\sqrt{\frac{2\pi kR(1/2-c_N)}{e^{2\pi kR(1/2-c_N)}-1}}~,
\ea
where $f^{(n)}_{\pm}$ are given by Eqs. (\ref{fsol}) and (\ref{majoranabc}) with $d_M=0$.

The above Eq. (\ref{matrixmaj}) can be written in terms of the mass eigenstates
\ba
 \frac12 m_{\chi_1}^{(0)}\chi_1^{(0)}\chi_1^{(0)}
+\sum_{n=1}^{\infty}\left[\frac12 m_{\chi_1}^{(n)}\chi_1^{(n)}\chi_1^{(n)} 
+\frac12 m_{\chi_2}^{(n)}\chi_2^{(n)}\chi_2^{(n)} \right]~,
\label{mass}
\ea
where $m_{\chi_1,\chi_2}^{(i)}$ are the mass eigenvalues and the corresponding mass 
eigenstates are denoted as $\chi_{1,2}^{(i)}$~\footnote{Note that, in the absence of 
Majorana mass term $d_M=0$, the apparent lepton number violation from the mass term
$m_{\chi_1}^{(n)}\chi_1^{(n)}\chi_1^{(n)}$ is canceled out 
by that from $ m_{\chi_2}^{(n)}\chi_2^{(n)}\chi_2^{(n)}$, so that the total lepton 
number is conserved as expected for 
$d_M=0$ \cite{ddg}.}.

For example, diagonalizing the above mass matrix by truncating at the first excited KK states in the parameter range $A_{00}\ll A_{01}\ll A_{11}$ of our interest (this hierarchy results from localizing a right-handed neutrino near the TeV brane in our scenario), we obtain the relevant approximate values of
the first few mass eigenvalues and the corresponding eigenstates
\ba
m_{\chi_1}^{(0)}&\sim& \left(A_{00}-\frac{A_{01}^2}{D_{N1}}\right)+\ldots,\qquad
m_{\chi_1,\chi_2}^{(1)}\sim D_{N1}\mp\frac{A_{11}}{2}+\ldots,\\
\chi_1^{(0)}&\sim&  N_+^{(0)}+\ldots,\qquad
\chi_{1,2}^{(1)}\sim \frac{1}{\sqrt2}[N_+^{(1)}\mp N_-^{(1)}]+\ldots,
\label{eigsta}
\ea
where higher-order contributions are at least suppressed by a factor of order 
${\cal O}(\frac{A_{01}}{A_{11}},\frac{A_{11}}{D_{N1}})\ll 1$ for the parameter range of our interest. 
Due to the Majorana mass term on the Planck brane, each KK excited state which originally consisted of degenerate even and odd states splits into a pair of nearly degenerate Majorana states.

The couplings of $\chi^{(n)}_{1,2}$ to the leptons and Higgs are obtained by substituting the Kaluza-Klein decomposition ansatz, which gives the following effective 4D Lagrangian
\ba
\label{4dlag}
 {\cal L}_{eff}\ni \int_{-\pi R}^{\pi R} dy  \left[\lambda_{\alpha \beta, \nu+}L_{\alpha+}(x,y) \tilde{H}(x,y)N_{\beta+}(x,y)+
 \lambda_{\alpha\beta,\nu-}L_{\alpha-}(x,y) \tilde{H}(x,y)N_{\beta-}(x,y) +h.c.\right]~,
\ea
where  $\lambda_{\alpha \beta,\nu\pm}$ are the 5D neutrino Yukawa couplings with mass dimension $-1/2$, and $\tilde{H}=i\sigma_2 H^*$. Note that there is no particular reason for the phases of $\lambda_{\alpha \beta,\nu \pm}$ to be aligned for different $Z_2$ parity $\pm$ or different flavor indices $(\alpha \beta)$, which will become important when we consider the CP asymmetry. We shall hereafter omit the flavor indices for brevity unless stated otherwise.

\section{Leptogenesis}
\label{lep}
We shall see how much baryon asymmetry can arise from the decay of KK excited states of right-handed neutrinos to obtain the analytical estimations for the parameter constraints. 

\subsection{CP asymmetry} 
In particular, consider the lepton asymmetry arising from the decay of the first KK excited states 
$\chi^{(1)}_{1,2}$. The Yukawa coupling terms relevant for $\chi^{(1)}_{1,2}$ decays are 
\ba
&&\sum_{n=0}^\infty\sum_{m=0}^\infty \bar{\lambda}_{\nu+}^{(1,n,m)}
L^{(n)}_+(x) \tilde{H}^{(m)}(x)N^{(1)}_+(x)
+
\sum_{n=1}^\infty\sum_{m=0}^\infty \bar{\lambda}_{\nu-}^{(1,n,m)}
L^{(n)}_-(x) \tilde{H}^{(m)}(x)N^{(1)}_-(x) +h.c.\\
&&=
\sum_{n}\sum_{m} \lambda_{\nu 1 \pm}^{(1,n,m)}
L^{(n)}_{\pm}(x) \tilde{H}^{(m)}(x)\chi^{(1)}_1(x)
+
\lambda_{\nu 2 \pm}^{(1,n,m)}
L^{(n)}_{\pm}(x) \tilde{H}^{(m)}(x)\chi^{(1)}_2(x)
+...~,
\ea
where $\bar{\lambda}_{\nu\pm}^{(1,n,m)}$ can be obtained from Eq. (\ref{4dlag}) by the integration over the fifth dimension 
\ba
\bar{\lambda}_{\nu \pm}^{(1,n,m)}&=&\frac{\lambda_{\nu\pm}}{(2\pi R)^{3/2}} \int^{\pi R}_{-\pi R} dy~f_{\nu \pm}^{(n)}(y)f_{H}^{(m)}(y)  f_{N \pm}^{(1)}(y)~,
\ea
and ${\lambda}_{\nu 1,2 \pm}^{(1,n,m)}$ can be obtained by substituting the mass eigenstates Eq. (\ref{eigsta}).
Depending on the mass parameter choice, the decay of $\chi^{(1)}_{1,2}$ to the excited KK states of leptons and Higgs can be kinematically forbidden. Note also that the $Z_2$ invariance prohibits the terms such as $L^{(0)}_+(x) \tilde{H}^{(0)}(x)N^{(1)}_-(x)$.

The CP asymmetry arises from the interference of the tree and one-loop diagrams of $\chi_1^{(1)}$ decay, when $\chi_2^{(1)}$ has a nearly degenerate mass. For the typical parameter range of our interest $m^{(1)}_{\chi_2}-m^{(1)}_{\chi_1}\gg \Gamma_{\chi_2^{(1)}}$,  it is estimated to be of order \cite{resold}
\begin{equation}
\label{cpn1}
\epsilon =\frac{\Gamma(\chi_1^{(1)}\rightarrow L^{} H^{\dagger})-\Gamma(\chi_1^{(1)}\rightarrow L^{c} H^{})}
{\Gamma(\chi_1^{(1)}\rightarrow L^{} H^{\dagger})+\Gamma(\chi_1^{(1)}\rightarrow L^{c} H^{})}
\sim \delta_{eff} \frac{ m_{\chi_1^{(1)}}}{\Delta m^2_{\chi^{(1)}}} \Gamma_{\chi_2^{(1)}}~,
\end{equation}
where
\begin{equation}
\delta_{eff}\equiv \frac{\mbox{Im}[({\lambda}_{\nu1}^{(1)*} {\lambda}_{\nu2}^{(1)})^2]}
{|{\lambda}_{\nu1}^{(1)}|^2 |{\lambda}_{\nu2}^{(1)}|^2}~,
\end{equation}
$\Delta m^2_{\chi^{(1)}}=(m^{(1)}_{\chi_2})^2-(m^{(1)}_{\chi_1})^2$ and the decay width is $\Gamma_{\chi^{(1)}_{1,2}}=|\lambda^{(1)}_{\nu 1,2}|^2 m_{\chi_{1,2}}^{(1)}/8\pi$ (with $\lambda^{(1)}_{\nu 1,2}$ indicating the dominant Yukawa couplings for the $\chi_{1,2}^{(1)}$ decay). Even though one may naively expect a very small CP asymmetry due to the tiny Yukawa coupling in our Dirac mass scenario, the small mass splitting together with the small wave function overlap partially compensates the tiny Yukawa couplings.

The CP asymmetry due to the out-of-equilibrium decay of $\chi_2^{(1)}$ can give the additional contributions of the same order as that of 
$\chi_1^{(1)}$ for a small mass splitting $m^{(1)}_{\chi_2}-m^{(1)}_{\chi_1}\ll m^{(1)}_{\chi_1}$ as one can see by switching $\chi_1\leftrightarrow \chi_2$ in Eq. (\ref{cpn1}).

Note also that the estimation in Eq. (\ref{cpn1}) takes into account only the self-energy contributions 
and there are also the vertex contributions 
which could give an additional CP asymmetry. We however shall not include it 
in the following analytical estimation 
for simplicity because the vertex contributions will not become much larger than self-energy contributions in our scenario, and in fact 
CP asymmetry of $\chi_1^{(1)}$ from the vertex corrections and that of $\chi_2^{(1)}$ cancel out as 
$m^{(1)}_{\chi_1}$ approaches $m^{(1)}_{\chi_2}$ in contrast to the self-energy contributions which contribute constructively \cite{pilareso,leptoextra}. We also note that the first
KK states alone can produce the sufficient baryon symmetry, while the
other higher KK states can contribute to the additional baryon asymmetry.
We shall hence use a simple analytical expression of Eq. (\ref{cpn1}) to estimate the order of magnitude for the baryon asymmetry.

Note that since $m^{(1)}_{\chi_{1}}\sim m^{(1)}_{\chi_{2}}$ and $\lambda^{(1)}_{\nu1}\sim \lambda^{(1)}_{\nu 2}$, we omit the subscripts $1,2$ for notational clarity in the following discussion when it does not cause any confusion.

\subsection{Baryon asymmetry}

\label{bau}
We assume a simple setup where $\chi^{(1)}$ decays out of equilibrium above the electroweak 
phase transition scale $T_c\sim {\cal O}(100)$ GeV so that the sphaleron effects are still active when it decays.  This requires the decay 
temperature $T_D\sim \sqrt{\Gamma_{\chi^{(1)}}}\sim \lambda^{(1)}_{\nu} \sqrt{m^{(1)}_{\chi}}$ 
($\Gamma$ is a decay rate) to be bigger than $T_c$ (for instance, for $m^{(1)}_{\chi}\sim$ TeV, 
this would require $\lambda^{(1)}_{\nu} \gtrsim 10^{-8}$). Letting the decay occur at the temperature below its mass scale 
$T_D\lesssim m^{(1)}_{\chi}$ (which gives $\lambda^{(1)}_{\nu}\lesssim 10^{-7}$ for $m^{(1)}_{\chi}\sim$ TeV), the out-of-equilibrium decay occurs for $\Gamma_{\chi^{(1)}}\lesssim H(T=m^{(1)}_{\chi})$ 
provided $\lambda^{(1)}_{\nu} \lesssim 10^{-7}$. 
Consequently, the resultant lepton asymmetry $Y_L$ due to the out-of equilibrium decay of  $\chi^{(1)}$ can be approximated by \cite{rocky} 
\ba
\label{chi1lepto}
Y_{L}\equiv\frac{n_L}{s}\sim  \frac{\epsilon n_{\chi^{(1)}}}{g_* n_{\gamma}}\sim  \frac{\epsilon}{g_*} ~,
\ea
where $s,n_{\gamma}$ are respectively entropy and photon density and $g_*\sim {\cal O}(100)$ is the number of relativistic degrees of freedom \footnote{We point out that the new degrees of freedom due to the KK excited states show up in four dimensions once the temperature cools down to the 
conformal symmetry breaking scale which is assumed to be around the TeV scale depending on the curvature of AdS or the location of the IR brane. 
Therefore, the relativistic degrees of freedom 
relevant for our discussions will not be changed greatly due to the degrees of freedom of the additional KK states.}. 
The sphaleron effects then convert it to the baryon asymmetry via the electroweak anomaly \cite{sphaleron} for the observed baryon asymmetry of the universe $Y_B\sim -\frac13 Y_{L}\sim  {\cal O}(10^{-10})$.

As a concrete example to illustrate that our scenario is a workable model, let us consider the decay of electron right-handed neutrinos. In this case, potentially stringent constraints arise from the electron Yukawa coupling, electron-neutrino Yukawa couplings for the zero and first KK modes and the baryon asymmetry of the Universe. These constraints can nevertheless be satisfied by adjusting the four main free parameters in our model, namely, the bulk mass parameters for the leptons and Higgs. For example, using $\pi kR\sim 34.5$ for $k e^{-\pi kR}\sim$ TeV, a bulk mass parameter choice of $(c_N,c_L,c_{e_R},b_H)\sim (-0.8,1.75,1.2,1.39)$ can realize $Y_B\sim {\cal O}(10^{-10})$ with the electron Yukawa coupling $\lambda_e^{(0)}\sim 10^{-6}$, electron-neutrino 
Yukawa coupling $\lambda_{\nu}^{(0)} \sim 10^{-19}$, the dominant Yukawa coupling for the first KK state $\lambda_{\nu}^{(1)} \sim 10^{-7}$ and 
the mass splitting of $(m^{(1)}_{\chi_2}-m^{(1)}_{\chi_1})/m^{(1)}_{\chi_1} \sim 10^{-8}d_M$  with the Planck scale Majorana mass parameter $d_M\sim 0.1$ and the 5D electron-neutrino Yukawa coupling of order $0.1$~\footnote{For our parameter choice, $m_{\chi}^{(1)}$ 
can decay to other first KK states such as $L^{(1)}_-$ and consequently a more precise quantitative analysis should take account of 
the effects of nonzero particle masses on the phase space suppression in the decay width and also on the sphaleron processes \cite{sphaleron, comp, masseffect}.}.

Note that the Majorana mass contribution to the zero mode right-handed neutrino becomes of order 
$A_{00}\sim 10^{-39}d_M$, and is indeed much smaller than the Dirac mass contribution. 
Hence our neutrino is still essentially Dirac in nature. We also point out that introducing
supersymmetry does not affect our TeV-scale leptogenesis mechanism. For 
instance, in the parameter range where a bulk Higgs is not quite peaked on the IR 
brane (such as in our concrete example), supersymmetry would be required to solve the 
usual gauge hierarchy problem. In fact, this supersymmetric  generalization would enlarge the
allowable parameter space and ease constraints on model building (for example,
the extra Yukawa coupling of a second Higgs doublet introduces an 
additional source of CP violation).

Of course to obtain a more precise quantitative estimate of the baryon asymmetry requires one to solve the Boltzmann equations which should take into account other non-trivial effects such as flavor, wash-out effects and thermal corrections \cite{comp,boltz}. Nevertheless our analysis shows that one can obtain an interesting viable model of the baryogenesis via leptogenesis scenario even if the Majorana 
mass is Dirac-like without invoking the see-saw mechanism, and where naturally small parameters are a characteristic feature of a warped extra dimension.

\section{Conclusion}
\label{dis}
We have presented a leptogenesis scenario where the neutrino masses are Dirac-like and yet the production of the baryon asymmetry is possible despite tiny neutrino Yukawa couplings. A key role was played by the small mass splitting in the nearly degenerate even and odd right-handed neutrino KK states. We emphasize that both the tiny Dirac Yukawa couplings and the small mass splittings naturally 
arise in the warped extra dimension.

Our model also has an interesting 4D dual interpretation via the AdS/CFT correspondence~
(see Ref.~\cite{fermass4} and references therein). The right-handed neutrino KK states are composite states in the dual gauge theory, while the left-handed neutrinos are elementary. Lepton number is a global symmetry of the gauge theory, but is explicitly broken in the elementary sector. The gauge theory couples to the elementary sector through irrelevant operators so only a small amount of lepton-number violation appears in the gauge theory. This corresponds to the small Majorana contributions to the right-handed neutrinos. The tiny Dirac couplings result from fermionic operators in the gauge theory with large anomalous dimensions. Therefore in the dual 4D theory it is the strong dynamics that is responsible for generating the small couplings and mass splittings.

While our mechanism is simple, other possibilities could also be incorporated in a warped dimension.
One of the most stringent constraints came from the condition that KK states should decay before the 
electroweak phase transition temperature $T_c$ is reached, after which the sphaleron effects are suppressed by a factor $e^{-T_c/T}$. This constraint 
however can be relaxed if the parameter tuning is possible such that $m^{(1)}_{\chi_2}
- m^{(1)}_{\chi_1 } \sim \Gamma_{\chi^{(1)}_{1,2}} $ which can lead to the 
resonant CP asymmetry becoming as large as of order unity $\epsilon \sim {\cal O}(1)$ to 
compensate such a suppressed sphaleron rate \cite{resold,pilareso,leptoextra}. 

Alternatively, if the Yukawa couplings are small enough to prevent the equilibration between the lepton 
asymmetry for the left and right-handed neutrino, the mechanism analogous to ``neutrinogenesis" 
can also be another possibility to produce the baryon asymmetry in the extra dimension scenarios 
\cite{hitoshi}. We also point out that even though we considered the mass splitting at tree level 
from the Majorana mass operator confined on the Planck brane, which can be justified from the 
natural lepton number violation via Planck scale physics, there can also be potential effects from radiative corrections as well \cite{leptoextra, carlos}~\footnote{In the parameter range of our interest, the radiative corrections arising from Yukawa and/or graviton couplings are suppressed by the additional small Yukawa couplings and/or small neutrino Majorana masses compared with the tree
level calculations, and therefore our tree level estimates of the mass splitting of first KK states would suffice for our discussions.}.

These additional effects and features deserve further study since the warped extra dimension 
provides an interesting alternative framework for baryon asymmetry production mechanisms at 
the TeV scale.

\subsection*{Acknowledgments}    
We thank J. Giedt, M. Shifman, K. Olive, L. Sevilla, M. Voloshin, M. Yamaguchi and T. Yanagida for useful discussions. The work of T.G. and K.K. was partially supported by DOE grant DE-FG02-94ER-40823 and that of M.Y. by JSPS Grant-in-Aid for Scientific Research No. 18740157 and the project of the Research Institute of Aoyama Gakuin University. T.G. is also supported in part
by an award from the Research Corporation.


\end{document}